\documentstyle[12pt]{article}

\renewcommand{\baselinestretch}{1}
\begin{document}
   \def\bea{\begin{eqnarray}}
   \def\eea{\end{eqnarray}}
\title{ Casimir energy for self-interacting scalar field in a spherical shell}
\author{$^{1,2,3}$
M.R. Setare  \footnote{E-mail: rezakord@yahoo.com}, $^{2,3}$R.
Mansouri \footnote{ E-mail:mansouri@sharif.edu }
\\
 {$^{1}$Department of Science, Physics group, Kurdistan University,
Sanandeg, Iran
 }\\
{$^{2}$  Department of Physics, Sharif University of Technology,
Tehran, Iran }\\{$^{3}$Institute for Theoretical Physics and
Mathematics, Tehran, Iran }}

\maketitle

\begin{abstract}
In this paper we calculate the Casimir energy for spherical shell
with massless self-interacting scalar filed which satisfying
Dirichlet boundary conditions on the shell. Using zeta function
regularization and heat kernel coefficients we obtain the
divergent contributions inside and outside of Casimir energy. The
effect of self-interacting term is similar with existing of mass
for filed. In this case some divergent part arises. Using the
renormalization  procedure of bag model we can cancel these
divergent parts.
\end{abstract}
\newpage
 \section{Introduction}
   The Casimir effect is one of the most interesting manifestations
  of nontrivial properties of the vacuum state in quantum field
  theory[1,2]. Since its first prediction by
  Casimir in 1948\cite{Casimir} this effect has been investigated for
  different fields having different boundary geometries[4-7]. The
  Casimir effect can be viewed as the polarization of
  vacuum by boundary conditions or geometry. Therefore, vacuum
  polarization induced by a gravitational field is also considered as
  Casimir effect.\\
   Casimir effect for spherical shells in the presence of the
   electromagnetic fields has been calculated several years ago\cite{{Boyer},{Schw}}. A
   recent simplifying account of it for the cases of electromagnetic and
   scalar fields with both Dirichelt and  Neumann boundary conditions on sphere
   is given in\cite{Nester}. The dependence of Casimir energy on the
   dimension of space for electromagnetic and scalar fields with Dirichlet boundary
   conditions in the presence of a spherical shell is discussed in\cite{{Milton},{Mil}}.
The Casimir energy for odd and even space dimensions and different
fields, including the spinor field, and all the possible boundary
conditions have been considered in\cite{Cog}. There it is
explicitly shown that although the Casimir energy for interior
and exterior of a spherical shell are both divergent, irrespective
of the number of space dimensions, the total Casimir energy of
the shell remains finite for the case of odd space dimensions.
More recently a new method have been developed for the scalar
Casimir stress on the D-dimensional sphere, in \cite{Sahar1}. In
this reference the regularized vacuum expectation values for the
scalar field energy-momentum tensor inside and outside a
spherical shell and in the region between two concentric spheres
have been calculated. Of some interest are cases where the field
is confined to the inside of a spherical shell. This is sometimes
called the bag boundary condition. The application of Casimir
effect to the
    bag model is considered for the case of massive scalar field
    \cite{bord} and the Dirac field \cite{Eli2}. We use the renormalization
    procedure in the above cases for our problem.\\
    Casimir effect for interacting fields has not been studied
    extensively. This is due to the lack of knowledge of true
    vacuum of an interacting quantum theory \cite{lang}.
   In the other hand the study of a massless scalar field with quartic
 self-interaction is very important in different subject of
 physics, for example in the Winberg-Salam model of weak
 interaction, fermions masses generation, in solid state physics \cite{{stan},{bel}},
 inflationary models \cite {guth}, solitons \cite{{dash},{raj}}
 and Casimir effect \cite{{lang},{jos}}.\\
In this paper we calculate the Casimir energy for spherical shell
with massless self-interacting scalar filed which satisfying
Dirichlet boundary conditions on the shell. We use the heat-kernel
techniquse to drive the zeta function for massless
self-interacting scalar field operator. Heat kernel coefficients
and zeta function of the Laplace operator on a D-dimensional ball
with different boundary conditions, both of them useful tools to
calculate Casimir energies, have been calculated in {\cite
{bek1}, \cite{bek2}}.The problem of calculating the determinant
of a Lplacian-like
 operator $A$ on a manifold $M$ is very important in mathematics
 \cite{ray} and physics\cite{{Remeo},{Elizalde},{bek2}}. In the cases which $A$ has a
 discrete spectrum the determinant of $A$ is generally divergent.
 The zeta function regularization is an appropriate way for these
 calculations. The zeta function method is a particular useful
 tool for the determination of effective action, where one-loop
 effective action is given by $\frac{1}{2}\ln det A$. Using the
 relation between zeta function and heat-kernel for operator $A$,
 one can find the zeta function.\\
 The paper is organized as follows: in the second section we briefly review the Casimir
  energy inside and outside of spherical shell in terms of zeta
  function. Then in section 3 we obtain the heat kernel coefficients
 for massless self-interacting scalar field inside and outside of
 spherical shell, after that we obtain the divergent part of
 Casimir energy and introduce the classical part of total energy
 of system, then using the bag model renormalization procedure
 \cite{{bord},{Eli2}} the renormalized Casimir energy can be
 obtained. Section is devoted to conclusion.

  \section{Casimir energy inside and outside of spherical shell  }
  We consider a massless self-interacting scalar field satisfying
  Dirichlet boundary conditions on a spherical shell in Minkowski
  space-time. The Casimir energy $E$ is the sum of Casimir
  energies $E_{in}$ and $E_{out}$ for inside and outside of the
  shell. The Casimir energies in-side and out-side of the shell
  are divergent individually. For free massless scalar field when
  we calculate the total Casimir energy, we add interior and
  exterior energies to each other. Now in odd space dimensions,
  divergent parts will cancel each other out \cite{Cog}.\\
  The Casimir energy in- and out-side of a spherical shell for
  massless free scalar field with Dirichlet boundary conditions is
  given by
   \begin{equation}
  E_{in}=\frac{1}{2R}(0.008873+\frac{0.001010}{\varepsilon}),
  \hspace{2cm}
  E_{out}=\frac{-1}{2R}(0.003234+\frac{0.001010}{\varepsilon}).
  \end{equation}
  where $R$ is a radiuse of spherical shell. Now we shall restrict
  our selves to a quartic self-interaction that is
  \begin{equation}
  V( \hat{\phi})=\frac{\lambda}{24}\hat{\phi}^{4}.
  \end{equation}
  We whish to consider the Casimir energy for this new case which
  given by
  \begin{equation}
  E_{0}=\frac{1}{2}\sum_{k}\lambda_{k}^{1/2},
  \end{equation}
 where $\lambda_{k}$ are determined by the eigenvalue equation
  \begin{equation}
(-\triangle +V''( \hat{\phi})) \phi_{k}(x)=\lambda_{k}\phi_{k}(x),
  \end{equation}
  which the scalar field satisfy Dirichlet boundary condition on
  the shell.\\
  Using the zeta function for operator $A$ which given by
 \begin{equation}
 A=\Box+v"(\phi).
  \end{equation}
  and
  \begin{equation}
\zeta_{A}(s)=\sum_{k}\lambda_{k}^{-s},
  \end{equation}
 we regularize $E_{0}$ by
  \begin{equation}
E_{reg}=\frac{1}{2}\zeta(s-1/2)\mu^{2s},
  \end{equation}
where we have introduced an unknown scale parameter $\mu$, whit
dimensions of mass to keep the zeta function dimensionless. The
zeta function (6) is the sum of zeta functions $\zeta_{in}(s)$
and $\zeta_{out}(s)$ for inside and outside of the shell,
therefore we can write
  \begin{equation}
E_{reg}^{in}=\frac{1}{2}\zeta_{in}(s-1/2)\mu^{2s},
\hspace{1cm}E_{reg}^{out}=\frac{1}{2}\zeta_{out}(s-1/2)\mu^{2s}.
   \end{equation}
 \section{Heat kernel coefficients inside and outside of spherical shell}
Now we use the heat-kernel techniquse to drive the zeta function,
the relation between zeta function and heat-kernel is given by
 \begin{equation}
\zeta _{A}(s)= \frac{1}{\Gamma (s)}\sum_{k}\int_{0}^{\infty} dt
t^{s-1}\exp(-\lambda_{k}t)=\frac{1}{\Gamma (s)}\int_{0}^{\infty}
dt t^{s-1}  K(t),
 \end{equation}
 where heat-kernel $K(x,x',t)$ satisfies the equation
\begin{equation}
(\frac{\partial}{\partial t}+ A)K(t,x,x')=0.
 \end{equation}
Now we write the short-time expansion of the heat-kernel $K(t)$
\begin{equation}
K(t)=\sum_{k}\exp(-\lambda_{k}t)\sim (4\pi t) ^{-3/2}
\sum_{k=0,1/2,1,...}^{\infty}(\int_{M}dv a_{k}+\int_{S^{2}}ds
c_{k})\exp(-tV''(\phi))t^{k}.
\end{equation}
The short-time expansion contain  both volume and a boundary part,
the coefficients whit integer numbers are volume parts and the
coefficients with half integer come from the boundary conditions,
for free field the heat kernel coefficients are as following
\begin{equation}
B_{k}=\int_{M}dv a_{k}+\int_{S^{2}}ds c_{k},
\end{equation}
where $M$ is inside and outside of spherical shell and $S^{2}$ is
the surface of spherical shell. The simplest first of $a_{k}$ and
$c_{k}$ coefficients for a manifold with boundary are given in
\cite{avra}, these coefficients in the presence of potential term
are given by
\begin{equation}
a_{0}=1,
\end{equation}
\begin{equation}
a_{1}=Q-\frac{1}{6}R
\end{equation}
 \begin{equation}
a_{2}=(Q-R/6) ^{2}-\frac{1}{3}\Box Q-\frac{1}{90}R_{\mu
\nu}R^{\mu \nu}+\frac{1}{90}R_{\mu \nu \alpha \beta}R^{\mu \nu
\alpha \beta}+\frac{1}{15}\Box R+\frac{1}{6}\tilde{R_{\mu
\nu}}\tilde{R^{\mu \nu}},
 \end{equation}
where $Q$ is a potential term, in our problem $Q$ is given by
 \begin{equation}
Q=V''( \hat{\phi})
 \end{equation}
 as one can see the $a_{k}$coefficients are functions of
  geometric quantities, $R_{\mu\nu\alpha\beta}$, $R_{\mu\nu}$ and
 $R$ are respectively, Rieman, Ricci and scalar curvature tensor.
 \begin{equation}
 \tilde{R}_{\mu\nu}=[ \nabla_{\mu}, \nabla_{\nu}],
 \end{equation}
where $\nabla_{\mu}$ is covariant derivative.
 Several first boundary coefficients in asymptotic expansion for
 Dirichlet boundary condition are as follow \cite {avra}
 \begin{equation}
c_{1/2}=-\frac{\sqrt{\pi}}{2},
 \end{equation}
 \begin{equation}
c_{1}=\frac{1}{3}K-\frac{1}{2}f^{(1)},
 \end{equation}
 \begin{equation}
 c_{3/2}=\frac{\sqrt{\pi}}{2}((\frac{-1}{6}\hat{R}-\frac{1}{4}R^{0}_{nn}+\frac{3}{32}K^{2}-
 \frac{1}{16}K_{ij}K^{ij}+Q)+\frac{5}{16}K
 f^{(1)}-\frac{1}{4}f^{(2)}),
 \end{equation}
where
\begin{equation}
f^{(1)}(z)=1/6+\frac{z^{2}}{6}(2+\frac{z^{2}}{2}-z(z^{2}+6)h(z)),
\end{equation}
\begin{equation}
f^{(2)}(z)=-1/6+\frac{z^{2}}{6}(-4+\frac{z^{2}}{2}-4z^{3}h(z)),
\end{equation}
\begin{equation}
h(z)=\int_{0}^{\infty}\exp -(x^{2}+2zx),
\end{equation}
 where $\hat{R}$
is the scalar curvature of the boundary, $K$ is trace of extrinsic
curvature tensor on the $S^{2}$,
\begin{equation}
K_{ij}=\nabla_{i}N_{j},
\end{equation}
where $N_{j}$ is outward unit normal vector.\\
The heat kernel coefficients $B_{k}$ for free scalar field with
Dirichlet boundary condition for interior region are given
by\cite{bek2}
\begin{equation}
B_{0}^{in}=\frac{4\pi}{3}R^{3},
\end{equation}
\begin{equation}
B_{1/2}^{in}=-2\pi^{3/2}R^{2},
\end{equation}
\begin{equation}
B_{1}^{in}=\frac{8\pi}{3}R,
\end{equation}
\begin{equation}
B_{3/2}^{in}=\frac{-\pi^{3/2}}{6},
\end{equation}
\begin{equation}
B_{2}^{in}=\frac{16\pi}{315 R}.
\end{equation}
The coefficients for exterior region are
\begin{equation}
B_{i}^{ext}=B_{i}^{in}, \hspace{2cm} i=1/2,3/2,...
\end{equation}
\begin{equation}
B_{i}^{ext}=-B_{i}^{in}, \hspace{2cm} i=0,1,2,...
\end{equation}
In the presence of self-interaction the heat kernel coefficients
for inside and outside  changed as following
\begin{equation}
B_{k}=(\int_{M}dv a_{k}+\int_{S^{2}}ds c_{k})(V''( \hat{\phi}))
^{3/2-k-s}
\end{equation}
then we have
\begin{equation}
B_{0}^{in}=\int_{M}dv
a_{0}V''^{3/2-s}=\int_{M}dv(\frac{\lambda}{2}\hat{\phi}^{2})^{3/2-s},
\end{equation}
\begin{equation}
B_{1/2}^{in}=\int_{S^{2}}ds
c_{1/2}V''^{1-s}=\frac{-\sqrt{\pi}}{2}\int_{S^{2}}(
\frac{\lambda}{2}\hat{\phi}^{2})^{1-s}ds,
\end{equation}
\begin{equation}
B_{1}^{in}=(\int_{M}dv a_{1}+\int_{S^{2}}ds
c_{1})V''^{1/2-s}=\int_{M}(
\frac{\lambda}{2}\hat{\phi}^{2})^{3/2-s}dv+\frac{2}{3R}\int_{S^{2}}
(\frac{\lambda}{2}\hat{\phi}^{2}) ^{1/2-s}ds
\end{equation}
\begin{equation}
B_{3/2}^{in}= \int_{S^{2}}ds c_{3/2}V''^{-s}
=\frac{-\pi^{3/2}}{6}\int_{S^{2}}ds
(\frac{\lambda}{2}\hat{\phi}^{2})^{-s}
+\frac{\sqrt{\pi}}{2}\int_{s^{2}}(\frac{\lambda}{2}\hat{\phi}^{2})^{1-s}ds,
\end{equation}
\bea B_{2}^{in}&=&(\int_{M}dv a_{2}+\int_{S^{2}}ds
c_{2})V''^{-1/2-s}=\int_{M}[\frac{\lambda^{2}}{4}\hat{\phi}^{4}\\
&-&
\frac{\lambda}{6}\Box\hat{\phi}^{2}](\frac{\lambda}{2}\hat{\phi}^{2})
^{-1/2-s}dv+\int_{S^{2}}\frac{4}{315 R^{3}}(
\frac{\lambda}{2}\hat{\phi}^{2})^{-1/2-s}ds,\nonumber
 \eea
 here $R$ is radiuse
of spherical shell. Now we can rewrite the zeta function (9) for
inside and out side of spherical shell.
\begin{equation}
 \zeta _{A}^{in}(s)=\frac{1}{(4\pi) ^{3/2}
\Gamma (s)}\sum_{k=0,1/2,1,...}B_{k}^{in}\Gamma(s+k-3/2)
\end{equation}

Similarly for outside we have
\begin{equation}
 \zeta _{A}^{out}(s)=\frac{1}{(4\pi) ^{3/2}
\Gamma (s)}\sum_{k=0,1/2,1,...}B_{k}^{out}\Gamma(s+k-3/2)
\end{equation}
The first five coefficients of heat kernel expansion contribute
to divergences of zeta functions inside and outside of spherical
shell. Using (8) we can write
 \bea
E_{div}^{in}&=&\frac{\mu^{2s}}{2 (4\pi) ^{3/2}\Gamma(s-1/2) }[
B_{0}\Gamma(s-2)+B_{1/2}\Gamma(s-3/2)\\
&+&
 B_{1}\Gamma(s-1)+
B_{3/2}\Gamma(s-3/2)+ B_{2}\Gamma(s)].\nonumber \eea
 Similarly
 \bea
E_{div}^{out}&=&\frac{\mu^{2s}}{2 (4\pi) ^{3/2}\Gamma(s-1/2) }[-
B_{0}\Gamma(s-2)+
B_{1/2}\Gamma(s-3/2)\\
&-&
 B_{1}\Gamma(s-1)+
B_{3/2}\Gamma(s-3/2)- B_{2}\Gamma(s)].\nonumber \eea
 We must
mention that for heat kernel coefficients $B_{k}$ in Eqs.(40,41)
$s\rightarrow s-1$.
 At this stage  we recall $E_{0}$ as given by (3) is only one
part of total energy. There is also a classical part. We can try
to absorb $E_{div}$ into the classical energy for inside and
outside of spherical shell separately. This technique of absorbing
an infinite quantity into a renormalized physical quantity is
familiar in quantum field theory and quantum field theory in
curved space \cite{Birrell}. Here, we use a procedure similar to
that of bag model \cite{{bord},{Eli2}}(to see application of this
renormalization procedure in Casimir effect problem refer to
\cite{{set1},{set2}}). The classical energy of spherical shell may
be written as,
\begin{equation}
E_{class}=PR^{3}+\sigma R^{2}+FR+K+\frac{h}{R},
\end{equation}
where $P$ is pressure, $\sigma$ is surface tension and $F$,
$K$,$h$ do not have special names. In odd space dimensions, for
free massless scalar field divergent parts inside and outside
cancel each other out, then we do not need to renormalize the
parameters in the classical energy, but in massive case, when we
add the interior and exterior energgies to each other, there are
only two contributions, which are divergent. In our problem there
is similar situation, as one can see, in Eqs.(40,41) when we add
divergent parts inside and outside, all terms cancel each other,
unless $B_{1/2}$ and $B_{3/2}$ terms \bea
E_{div}&=&E_{div}^{in}+E_{div}^{out}=\frac{\mu^{2s}}{(4\pi)
^{3/2}\Gamma(s-1/2)}[B_{1/2}\Gamma(s-3/2)\\
&+&B_{3/2}\Gamma(s-1/2)].\nonumber
 \eea
Therefore the Casimir energy for this general case becomes
divergent. The total energy of the shell maybe written as
\begin{equation}
E^{tot}=E_{0}+E_{class}
\end{equation}
In order to obtain a well defined result for the total energy, we
have to renormalize some parameters of classical energy according
to the below:
\begin{equation}
K\rightarrow K-\frac{\mu^{2s}}{(4\pi)
^{3/2}}[\frac{-\pi^{3/2}}{6}\int_{S^{2}}ds
(\frac{\lambda}{2}\hat{\phi}^{2})^{-s}
+\frac{\sqrt{\pi}}{2}\int_{s^{2}}(\frac{\lambda}{2}\hat{\phi}^{2})^{1-s}ds]
\end{equation}
\begin{equation}
\sigma \rightarrow \sigma+\frac{\mu^{2s}}{16 \pi R^{2}
\Gamma(s-1/2)}\Gamma(s-3/2)[\frac{-\sqrt{\pi}}{2}\int_{S^{2}}(
\frac{\lambda}{2}\hat{\phi}^{2})^{1-s}ds].
\end{equation}
Hence the effect of the self-interacting scalar quantum field is
to change, or renormalize some parameters of classical energy of
system. Once, the terms $E_{div}$ have been removed from $E_{0}$,
the remainder is finite and will be called the renormalized
Casimir energy. Using Eq.(8) and Eqs.(38,39) we have
 \bea
E_{ren}&=&E_{0}-E_{div}=\frac{\mu^{2s}}{2}(
\zeta_{A}^{in}(s-1/2)+\zeta_{A}^{out}(s-1/2))\\
&-&E_{div}=\frac{\mu^{2S}}{(4\pi)
^{3/2}\Gamma(s-1/2)}\sum_{k=5/2,3,...}\Gamma(s+k-2)
(
B_{k}^{in}+B_{k}^{out}).\nonumber \eea

    \section{Conclusion}
In this paper we have considered the Casimir energy for massless
self-interacting scalar field in a spherical shell. The scalar
field satisfy Dirichlet boundary condition on the shell. Unlike to
the main part of previous studies on the scalar Casimir effect,
here we adopt the Casimir energy with interacting quantum field.
To obtain the divergent parts of energy inside and outside of
spherical shell, we calculate the first five  heat kernel
coefficients for operator $A=-\Box+V''( \hat{\phi})$. Previous
result of heat kernel coefficients for Laplace operator
\cite{{bord},{bek2}} have been changed for our problem. The new
result are given by Eqs.(32-37). In free massless case when we add
divergent parts inside and outside, all divergences cancel each
other out, but in interacting scalar field case, exactly similar
to the massive case, divergent parts inside do not cancel the
divergent parts outside of shell. The renormalization procedure
which is necessary to apply in this situation is similar to that
of the bag model \cite{{bord},{Eli2}}. At this stage we introduce
the classical energy and try to absorb divergent part into this
classical energy. The renormalization can be achived now by
shifting some parameter of classical energy by an amount which
cancel the divergent contribution. After the subtraction of
divergent contribution, the remainder is finite and will be called
the renormalized Casimir energy.

\end{document}